# AI-Powered GUI Attack and Its Defensive Methods

Ning Yu, Zachary Tuttle, Carl Jake Thurnau, Emmanuel Mireku
nyu@brockport.edu,ztutt2@brockport.edu,cthur3@brockport.edu,emire1@brockport.edu
State University of New York The College at Brockport
Brockport, New York

## ABSTRACT

Since the first Graphical User Interface (GUI) prototype was invented in the 1970s, GUI systems have been deployed into various personal computer systems and server platforms. Recently, with the development of artificial intelligence (AI) technology, malicious malware powered by AI is emerging as a potential threat to GUI systems. This type of AI-based cybersecurity attack, targeting at GUI systems, is explored in this paper. It is twofold: (1) A malware is designed to attack the existing GUI system by using AI-based object recognition techniques. (2) Its defensive methods are discovered by generating adversarial examples and other methods to alleviate the threats from the intelligent GUI attack. The results have shown that a generic GUI attack can be implemented and performed in a simple way based on current AI techniques and its countermeasures are temporary but effective to mitigate the threats of GUI attack so far.

## 1 INTRODUCTION

Cybersecurity is a history of arms race, where a constantly evolving cat-and-mouse game is played by attackers and defenders. The emerging computing technology of every new era has fueled attackers with new capabilities as well as an amount of vulnerabilities to implement their nefarious plots. When we are on the cusp of a new era: the artificial intelligence (AI) era, the shift to machine learning and AI will be the major progression in the next generation of cybersecurity. In recent years, there is an increasing trend in using artificial intelligence, machine learning, data mining, and deep learning techniques in static and dynamic malware analysis, and anomaly detection [5, 24]. It is predictable that the AI technology will be used in more of greater importance in the near future. Even though, currently AI-based cybersecurity research is still in its infancy. In 2018, IBM researcher demonstrated the AI-based attacks and indicated the coming of next-generation cybersecurity threats [9]. Similar AI-based cybersecurity attacks [12, 21] were illustrated that the email compromise can occur by exploiting on machine learning and artificial intelligence. A neural network was the commonly used hacking tool that can learn how to break into web applications [15]. Actually, these aforementioned techniques take advantage of the vulnerabilities of the existing system. However, none of them are related to AI-based GUI attack.

From the perspective of computer security, any human's habit /favorite in a computer system can be tracked by machine learning and artificial intelligence because such habits /preferences are regarded as specific patterns susceptible to malicious learning machines. Moreover, unfortunately such personal data are often open and unprotected. For example, when one uses his/her computer, the GUI is exposed. In case the computer was infected by AI-based malware, the personalized desktop information can be perceived and collected by the malware automatically. The article aims to demonstrate such an attack prototype where the victim's desktop can be targeted by stealthy malware and the GUI properties can be exploited and recognized by the malware to perform a series of malicious attacks. As a typical example, the malicious malware could be able to stealthily identify what the user is operating over a bank website and it could launch an attack to automatically transfer money from a bank account to another account when the login buttons on the webpage were identified and a click event was triggered by the malware. In the current software configuration, the GUI software have no immunization from such a type of attack.

Therefore, on the other side, the security community needs secured Apps or GUIs that are expected to be immune from such AI-based attacks. We cannot simply wait to start preparing our defenses until the attacks are found in the wild. To that effect, the AI-based attacks and new traits need to be identified and compared to traditional attacks. It would be beneficial to focus on monitoring and analyzing how AI-based attacks behave through user's devices, and activating events when the user is taking routine actions. This article means to help identify this type of attack. Moreover, the future software engineering should have the feature that could prevent from automated AI-based attacks. The design of such a defensive method essentially is about to discover the vulnerabilities of current AI system. Adopting adversarial examples against data-driven AI systems was a novel attacking scenario targeting at AI vulnerabilities [19]. Thus, it also fits to prevent the attack from the AI-based malware. Adversarial examples are small perturbations in the input image to victim machine learning models where an attacker has intentionally crafted the input data to trigger the machine's recognition mistake/error while a human can still easily recognize the image without visionary impacts.

The rest of this article is organized as follows. Section 2 introduces the latest techniques on the two sides of the arm race in AI-powered cybersecurity. Section 3 discusses the approach to perform a GUI attack. Section 4 illustrates the effectiveness of the GUI attack. Sections 5 and 6 gives several methods to defend the GUI attack and their performance. At last, Section 7 concludes the article and makes a discussion about the future work.

## 2 RELATED WORK

In 2017 DeepHack the open-source hacking AI learned how to break into web applications using a neural network, which can exploit multiple kinds of vulnerabilities, opening the door for a host of AI-powered hacking systems in the future [15]. In 2018 one of the most recent and advanced examples of AI-powered malware called DeepLocker was first demonstrated by IBM researchers [9]. The two breakthroughs have indicated that an arms race has been started between those who wish to attack and those who wish to defend neural networks with advancements in technology and a resurgence in artificial intelligence and machine learning.

On the one hand, the benign-looking malware can stealthily and automatically collect and analyze the attributes of targeted subjects with normal looking and behavior so that pattern based anti-malware software is not eligible of finding them. These malware such as DeepLocker also can provide a concealment method for the cybersecurity attack based on the deep neural network to generate a key according to user's attributes [9]. In recent years, there has been an influx of research into artificial intelligence and neural networks on cybersecurity. As typical examples, researchers used AI and deep learning on intrusion detection over traditional network traffic [1, 8, 18, 24] and even in-vehicle networks [22]. Rege and Mbah summarized the applications that Machine Learning made to cyber crime, including several aspects: Spear Phishing, Unauthorized Access, and Evasive Malware [16]. Crafting email using AI techniques was illustrated in 2015 by Palka et al. [12]. A business email compromise attack was demonstrate in 2017 to show a technical exploits on machine learning and artificial intelligence [21]. Zhang et al. proposed a stealthy attacking method targeting Voice Assistant on smart phones by using an Intelligent Environment Detection module to hide the attack from being noticed by users and choose an optimal attacking time [27]. As a commonly used technique, the neural networks are not only used to predict and detect the network intrusion through KDD traffic dataset [18, 24] but also utilized to generate cryptographic keys according to user's attributes illustrated by IBM DeepLocker. Due to the highly complex nature of the neural network, even if the malware had been found, it would be very difficult for malware analysts to determine who or what the malware was searching for. Without knowing these things, it would be impossible to decipher the trigger condition, meaning that the payload would never be unlocked and remain unable to be studied [9]. Meanwhile, the open-source AIs have been widely used by hacker, further lowering the hacking cost. DeepHack used the open-source hacking AI to learn how to break into web applications [15]. In the recent literature of 2019 [23], researchers discovered that considerable information on companies and individuals can be easily gained by attackers by using open source intelligence that increases the threat of targeted attacks. Open-source libraries were also employed to conduct a low-cost attack on cyber authentication system [19, 25].

On the other hand, much work has been done in the space of adversarial examples, proposing and classifying various types of attacks such as black-attacks, in which the attacker does not know anything about the structure of the network [20] as well as white-box attacks in which the attacker has some knowledge about the structure of the neural network. It has also been found that adversarial examples are transferrable among almost any artificial intelligence model. Shukla et al. first found that adversarial examples were transferable between the same neural network being trained on different data sets [20]. Meanwhile, Papernot et al. found that an adversarial example that effectively fools one model will most likely fool a completely separate model in the same manner or fashion [13]. This is critical for black-box attacks as an attacker can use a substitute model to generate adversarial examples and then transfer them to the model in which they are invoking the attack. Yuan et al. explored possible defenses and countermeasures that could be employed against adversarial examples. Bayesian classifiers and Gaussian Processes were used to propose new neural networks called Gaussian Process Hybrid Deep Neural Networks [26]. These networks were found to be comparable to general deep neural networks in performance, yet more robust against adversarial examples [26].

Since the essence of adversarial examples is an attack against the blind-spot of the neural networks, the improvement in the robustness of the neural network is the countermeasure against such an attack. Neural network robustness can be evaluated in a few different ways. One such way is to attempt to prove a lower bound, or generate attacks that can demonstrate an upper bound [4]. The former method, while effective, is much more complex to implement in the physical world. With the latter, if the attacks used on the neural network are not strong enough, an upper bound may not prove useful [4]. Another powerful method to increase robustness of a neural network is defensive distillation. Defensive distillation is motivated by the goal to reduce the size of either Deep Neural Network (DNN) architectures or ensembles of DNN architectures [14]. Doing this allows for the reduction of computing resource needs and in turn allows for deployment on resource constrained devices such as a smart phone [14]. This technique is done by extracting the class probability vectors that are produced by a DNN and transfer them to a second DNN of reduced dimensionality during training without observing a loss of accuracy [14]. Furthermore, Carlini and Wagner claimed that defensive distillation was no more resistant to targeted misclassification adversarial examples than an unprotected neural network and they have showed that with slight changes to the attack model, they could achieve successful targeted misclassification [3].

## 3 APPROACH OF GUI ATTACK

### 3.1 Malware Design

The main objective of the malware was to detect, recognize, and locate icons on the victim's GUI such as desktop and tray/task bar,

specifically web browsers, and then signal the click event to these icons so that the login information stored by the victim in the web browser can be taken advantage of by the malicious malware to steal the victim's asset. For the purpose of prototype demonstration, we choose to stealthily log into the victim's Blackboard account as our goal. The targets are Windows users who have either one of the Google Chrome, Mozilla Firefox, Microsoft Edge, Opera browsers or all with their user login information saved into the user name and password field respectively. The more browsers the user has the better the attack would be; if the user has more than one browser then the chances of detecting the browser would be high. The malware will execute successfully if the the victim of the malware has internet connection, at least one of the browsers, their information saved in their blackboard login web page on Microsoft Windows 10.

The malware may be implemented in two ways: one is based on AI-powered object detection model and another is implemented on non-AI based method. The malware is specifically made to detect all the four browsers. With this ability, the detection of a browser on a user's computer desktop and tray/taskbar becomes an easy and efficient task. Once an icon is detected, move to the location, open the detected browser and execute the attack by logging into the victim's Blackboard account.

If a browser was not detected on the tray/task bar, the malware may see the possibility that the user's desktop might have some of the browsers after all windows were minimized. After doing this and there was no detection then the start menu would be opened and the link would be entered straight into the start menu search box and searched. When the computer opened the link with its default browser, then it logged in the user and execute the attack. The work flow of the malware is demonstrated in Figure1.

## 3.2 Object Recognition Model

It has shown on Figure1 that the object detection is located at the center of the malware, determining the success of an attack. TensorFlow Object Detection (TOD) model is used as the component recognition in our framework to detect icons and perceive their locations. Open-source TensorFlow detection model repository offers a magnitude of models, able to be trained further and integrated into image recognition work, in which faster regional Recurrent-CNN inception [17] was scored high in both accuracy and reasonable computation speed. Thus, after integrating the pre-built model to our detection program, we trained the model and extracted location information for feeding the coordinates to our malware program, which ensures high accuracy in object detection.

The TOD detection model used on top of our recognition model was built using Google object detection directory within the TensorFlow repository. Figure 2 illustrates the loss over a 10,000-iteration period for training this detection model. It shows how quickly each of the icons individually approached an optimal loss value as well as the loss value of training all icons together. The main loss graph of interest can be found at the top left identified as the classification loss. This loss determines the overall accuracy of how well the model performs given how low the loss value is after a specified number of iterations. The ideal loss value to reach was at or below, 0.05. In almost all of the Loss Graphs, the loss value quickly approached 0.05 at around 4 to 5 thousand interactions. Due to the training script that was used however, checkpoints were only saved every 10 min, and about 3k iterations were accomplished in those 10 minutes. So for the model, the ideal inference graph to be exported for classification was around 5 to 6 thousand interactions. If the Inference graph was exported at 3 thousand, the model might not have been as accurate as it could be because it could be trained to reach a loss value lower than 0.05. However if one exports the inference graph of the checkpoint that was saved at 8 to 9 thousand iterations, one runs the risk of running a model that wrongly classifies objects due to overfitting. In Figure 2, classification loss is the loss between the detection and a set of icons; localization loss is the loss of wrapping box around the character and its real location; RPN loss is the loss of combining the previous classification and localization losses; objectness loss is the loss of classification based on the bounding boxes; total loss is the total loss between classification, localization, RPN localization, and objectness losses; clone loss is the same as the total loss.

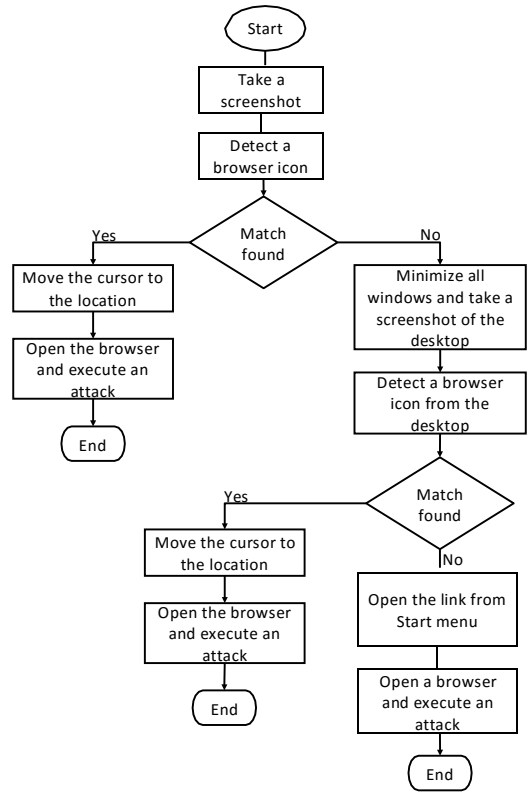

**Figure 1: The Work Flow of GUI Malware**

## 3.3 Data Acquiring and Labeling

In order to acquire data for training purposes, a python data creation script was made that helped facilitate this process. Approximately 200 images were captured per each icon for basic training in order to acquire the ability to perceive the icon on the screen. 100 of those images were trained based on the much smaller tray icon, and the

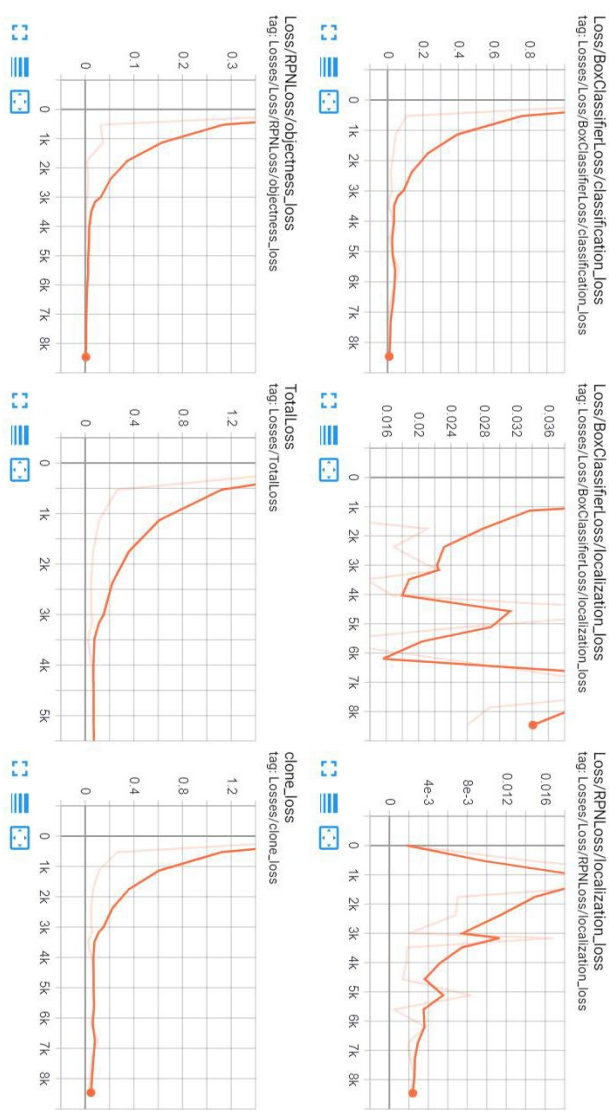

Figure 2: Loss Over Training Iterations

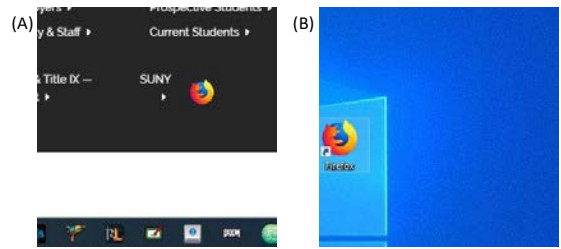

**Figure 3: (A) 300 by 300 cropped tray icon image and (B) 300 by 300 cropped Desktop shortcut image**

After properly segmenting the data into the correct folder, a script is run to convert all of the .xml files to their corresponding test and train .csv files. This is necessary to create TensorFlow records corresponding to the test and train data, which is a converted binary file format that helps increase the performance of the model. Meanwhile, another script is used to generate these TensorFlow records before the model is ready to be trained.

## 4 EFFECTIVENESS OF GUI ATTACK

The performance of the GUI attack depended on the predictive accuracy of the object detection model. Thus, first we evaluated the performance of the recognition model itself. Then, we compared the performance of two implementations for the GUI attack: one was implemented by AI model and another was implemented by non-AI model.

### 4.1 Tests of Object Detection Model

Four tests were uniquely generated with screenshots that contained all possible detected objects. The final average test combined all four tests to show the overall performance of the tests. Two environments were tested on, detecting icons that were displayed in the system tray at the bottom of the window as well as the Desktop environment, as shown in Figure 4. The Tray and Desktop environments differ on two key fundamental factors. The first factor is size, the tray icons are noticeably smaller so detection will be harder for the tray icons in comparison. The second factor is the fact that the icon shortcuts on the Desktop environment contain a small arrow indicating that it is a shortcut. For this reason, half of the training data consisted of shortcut icons and half of the training consisted of the tray sized icons to allow for enough varied data for proper detection.

Looking at each icon independently, Chrome and Firefox seemed to be the most consistently detected icons in comparison to the Edge and Opera icon. One reason why this could be, would be due to the simplistic nature of the Edge and Opera Icon. The Edge icon at its basic level is a blue E, while the Opera icon is just a red O. For this reason, misdetection was higher when testing different screenshot environments. In Test 3 in Figure 4, the chrome icon was misdetected as the opera icon at 75% as denoted as **(O)**.

The model tests have shown that the confidence level of recognition is very high with the average more than 95% in both desktop and tray icons except the Chrome in the tray, in which we calculated the average excluding Test 3 denoted as **\***. The recognition other 100 images were trained based on the larger desktop shortcut icon. Meanwhile, a transparent .png of the icon were placed at the exact cursor position. Once the transparent .png was masked onto the image, a 300x300 crop is immediately generated around the transparent image as the training data set.

After all of the data are acquired, it is needed to label for the supervised learning. The labeling was done using an open source program called Labelimg. This program helps generate an .xml file with the exact coordinates of the bounding box placed around the icon. The name of the .xml file has the exact name as the .png version. When all of the images are properly labeled, 20% of the images are placed in a test folder and the rest of the 80% are kept in a training folder.

ability of the TOD model provides the strong support for the GUI malware to perform an effective and accurate attack against GUI systems.

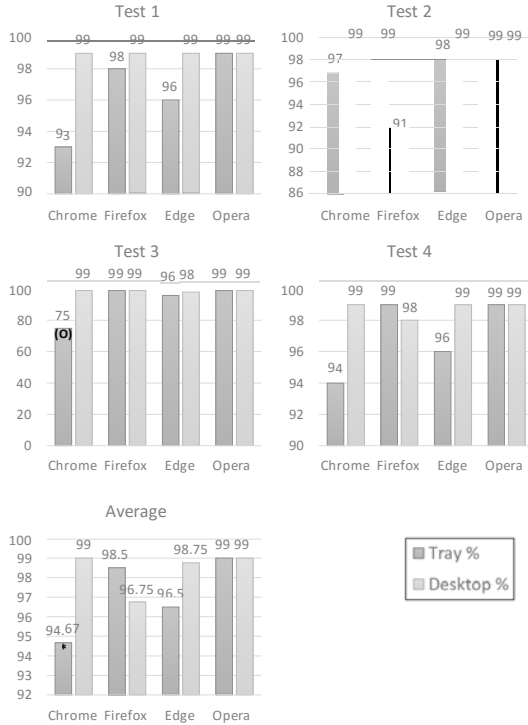

Figure 4: Results of the Model Tests. (o) denotes it misclassified as Opera. * denotes the average taken by Tests 1,2 & 4.

## 4.2 Non-AI Approach vs. AI-Powered Approach

Python libraries such as OpenCV, YOLO, Single shot detectors and many more were investigated by us. However, the OpenCV library is chosen as the alternative non-AI method to compare with our AI-based object recognition method, which is based on the following knowledge. (1) It is known as Open Source Computer Vision and Machine Learning Library. (2) It was built to provide a common infrastructure for a computer vision application and accelerate the use of machine perception in the commercial products. (3) It also makes it easier for businesses to utilize and modify the code because of it being a BSD-license product. The template matching, a non-AI method from OpenCV library, is used for comparison to our AI-based method. Template matching is a technique for finding areas of an image that match (are similar) to a template image. This method takes two inputs which are the source image (the image containing the image to be detected) and the template image (the image that will be detected in the source image). The goal of this method is to detect the highest matching area. So, in order the for the method to find this, it compares the two images by sliding the template image against the image source; the template image is moved at one pixel at a time from left to right and up to down. At each location that template is slid across, a metric is calculated so it gives a result which will yield how accurate or similar the area is compared to the template image.

The comparison between non-AI method and AI-based method is demonstrated on Figure 5. The AI-based malware was executed and validated its effectiveness that make the GUI attack work successfully with high accuracy since it used trained data/images sets for the detection. On the contrary, Template Matching, the non-AI method is not very accurate and may not detect all the images you want it to detect. It might also detect something that is not right. With all the optimization, the non-AI malware failed to detect some of the images and the detection value of the image is not very high. From the template matching data set result, it can be concluded that the detection are very inconsistent event though it detected the icons; it sometimes would detect the image at a higher percentage and sometimes it would not at all.

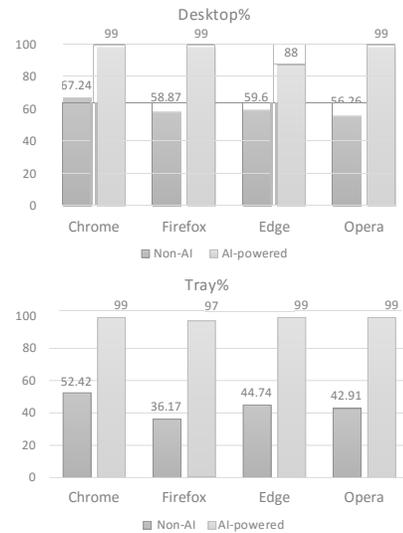

Figure 5: Non-AI Method (template match) Vs. AI-powered Method in Recognition Accuracy of GUI Attack

## 5 DEFENSIVE METHODS
### 5.1 Fast Gradient Sign Method

According to the literature [19] published in Jan 2019, it was an innovative way to take advantage of adversarial examples to spoof AI-based attacks. Adversarial examples are forged inputs to a neural network, resulting in an incorrect output from the network. In order to find the adversarial examples that can lead to the misclassification in a malicious learning machine, the approximate search algorithm such as the Fast Gradient Sign Method (FGSM) [10] was adopted in the research.

A perfect AI system is expected to be capable of predicting the correct class $y$ of any given input $x$. In an AI-powered network intrusion detection system, the input $x$ represents the features of the network packet, such as protocol, service, destination, source, connection, and so on. The class $y$ indicates the class of attack, which can be a binary value or multiple classes depending on the requirement. For an ideal AI system, the expectation of the model error $e$ between the prediction $y_1$ and the truth $y_0$ is optimized to the minimum.

In an AI system, internal architectures are hard to access but data can be available. Assuming that the same training data can be shared by attackers and the AI system, an adversarial example $x^*$ is defined by perturbing an originally correctly classified input $x$. Finding $x^*$ is reciprocal with a constrained optimization problem. The adversarial sample $x^*$ causes the most expected loss (prediction error) to reflect the classification error that is subject to a constraint on the minimum deviation from the original input $x$. That is, the smallest modification causes a misclassification $x^* = x + ar\partial min\{|z|: f(x+z) = t\}$, where $z$ is the perturbation of input, $|z|$ is the norm of perturbation, and $t$ is the target class different from $f(x)$ [7].

In order to find the adversarial examples that can lead to the misclassification, the adopted Gradient-based approximate search algorithms in this project include Limited-memory Broyden–Fletcher–Goldfarb–Shanno (L-BFGS), the Fast Gradient Sign Method (FGSM), and the Jacobian Saliency Map Approach (JSMA).

The L-BFGS method was used for computer-vision model to misclassify an image where an imperceptible non-random perturbation was added. L-BFGS can be revised by using the Adam optimizer to reduce the size of the perturbation [4]. The downside of both methods is the high computing cost. However, it is hard to make a trade-off between computing cost and effectiveness. In practice, it is reasonable to modify L-BFGS by simply running the L-BFGS in less iterations to have a less expensive cost at a lower accuracy.

Compared with L-BFGS, FGSM is an approach with a lower computing cost [10]. FGSM can maximize the prediction error and keep the size of input perturbation. The adversarial example can be formulated as $x^* = \max J_f(x)$ subject to $\|x^* - x\|_\infty$ where $J_f$ is the way to measure the expected loss (prediction error). By substituting $J_f$ with a first-order Taylor series approximation and taking the gradient at $x$, the equation is further formulated as $x^* = x + \epsilon sign(\nabla_x J_f(x, y))$, where matrix $J_f = \left[\frac{\partial f_t}{\partial x_i}\right]_{ij}$ is the Jacobian matrix. The Taylor series expansion uses the linear approximation and incurs a lower running time.

## 5.2 Gaussian and Salt-and-Pepper Noise

The original image is still easily recognizable to a human after Gaussian noise is added. Since Gaussian noise is a form of additive noise, in which each pixel in the noisy image is the sum of the true value of each pixel and a random, Gaussian distributed value. The noisy image is described by $A(x,y) = H(x,y) + B(x,y)$ where $A(x,y)$ is the function of the noisy image, $H(x,y)$ is the function of image noise, and $B(x,y)$ is the function of the original image [2]. The Gaussian noise we added to the chrome icon had a mean value of 0 and a sigma value of 20.0.

In addition to this, we also tested a slightly lower standard deviation value of 15. Another form of noise we tested in our research is Salt-and-Pepper noise. Salt-and-Pepper noise is also commonly known as impulse valued noise or data drop noise due to the statistical drop of original pixel values [9]. It results in sharp and sudden disturbances in the image. To add Salt-and-Pepper noise to an image, we calculate the noise for a given pixel $(i, j, k)$ as Noise $(i, j, k; p)$ where $i(i, j, k)$ represents the original, intact pixel value and $p$ represents the noise density value [11]. With Salt-and-Pepper noise, we have progressively dark pixels present in bright regions of the picture and vice versa [9].

## 5.3 Poisson and Speckle Noise

Our third chosen technique of noise addition was Poisson, also called Photon noise. Appearance of this noise is often the result of the statistical nature of many electromagnetic waves such as x-rays, gamma rays, and visible light [9]. Sources have random fluctuations of photons resulting in an image with spatial and temporal randomness [9]. Finally, the final image technique we worked with was the addition of Speckle noise, also known as multiplicative noise. The probability density function of Speckle noise follows a gamma distribution [2] and this type of noise can exist in an image similar to Gaussian noise [9].

## 6 PERFORMANCE OF COUNTERMEASURES

Figure 6 shows the results of several countermeasures against the GUI attack. The adversarial examples created through FGSM was effective in preventing the neural network from recognizing our web browser icons. For example, in our tests using Google Chrome, with the resize sub-method, the image classifier was unable to recognize the icon. The resize method is to enlarge the object first to feed the TOD model since it is so small in size and shrink the enlarged image into the original size after the FGSM operation. Alternatively, the pad and crop method is to pad the original icon with background color and to crop the object only after the FGSM operation.

In the case of the pad and crop method, the classifier was able to recognize the chrome icon with 77% confidence. Upon testing our methods with Microsoft Edge, with the pad and crop method, we had a result of 67% confidence in recognition. Using the resize method resulted in our image classifier being unable to recognize the Edge icon. The resize method on Firefox gave us a result of the network being unable to recognize the icon while the pad and crop method provided approximately 75-80% confidence. Finally, in our testing of the Opera icon, were found approximately 65% recognition confidence while using the pad and crop method. For the resize method, we found that the Opera icon was recognized sometimes with a confidence of around 75-80%, and in rare cases, it was recognized as Microsoft Edge with around 77% confidence. As a result of our testing using FGSM, we were in most cases able to drastically lower the ability of our neural network classifier to identify web browser icons. In each of these tests, once the perturbed icon was pasted into the original input image, the icon on the background image was still very recognizable to human eyes, which meets the main criteria of our experiment.

As for Gaussian noise added to the original images, we were able to lower the classifier's confidence level in recognition by differing amounts. The amount of drop-in confidence level depended on how much noise we wanted to add to the image. For Chrome, we needed a bit more Gaussian noise than the other icons to drop the confidence of recognition by the same levels. Using a distribution

| Original | FGSM + Resize | FGSM + Pad&Crop | Gaussian | | Salt-Pepper | Poisson *** | | Speckle ***** |
|---|---|---|---|---|---|---|---|---|
| 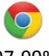 97-99% | 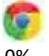 0% | 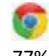 77% | 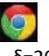 δ=20 81-85% | 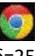 δ=25 0% | 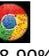 88-90% | 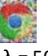 λ = 500 0% | 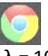 λ = 100 88% | 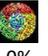 0% |
| 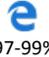 97-99% | 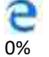 0% | 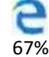 67% | 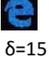 δ=15 82% | 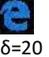 δ=20 0% | 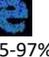 95-97% | 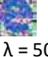 λ = 500 0% | 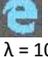 λ = 100 96% | 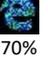 70% |
| 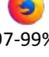 97-99% | 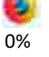 0% | 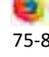 75-80% | 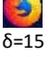 δ=15 86% | 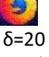 δ=20 75% | 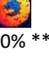 80% ** | 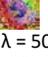 λ = 500 0% | 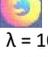 λ = 100 92% | 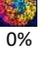 0% |
| 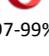 97-99% | 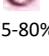 75-80% * | 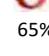 65% | 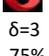 δ=3 75% | 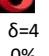 δ=4 0% | 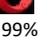 99% | 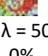 λ = 500 0% | 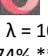 λ = 100 74% **** | 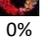 0% |

Note that the percentage means the confidence level of recognition provided by the AI-powered object recognition method.
\* In rare cases, while using this method on Opera, the icon was recognized as Microsoft Edge with around 77% confidence
\*\* Firefox required more noise to drop recognition level, icon became much less recognizable from original image
\*\*\* At λ = 100, icon colors were often inverted once put through the image classifier
\*\*\*\* For Opera at λ = 100, icon was recognized as Google Chrome with 74% confidence
\*\*\*\*\* Icons were mostly overpowered by speckle noise, making the method ineffective for our theory

**Figure 6: Countermeasures of GUI Attack**

with a mean of 0 and a standard deviation of 20, the image was still easily recognizable by a human while the icon was recognized by the image classifier with a confidence level of around 81-85%. The noised image with a standard deviation of 25 cannot be recognized by the recognition model. It seems that the more noise added in it, the harder the image classifier was able to recognize it. However, it is harder for human eyes to recognize it. For Microsoft Edge, a standard deviation of 15 was enough to bring confidence level of recognition to 82%, followed by the icon being unrecognizable at a standard deviation of 20. Next, for Mozilla Firefox, using a standard deviation of 15 dropped confidence level of recognition to 86%, and a standard deviation of 20 bringing it further down to 75%. Finally, in the case of the Opera icon, there were some unique results. The standard deviation of noise required to lower the confidence level of recognition to 75% was only 3. At a standard deviation of 4, the neural network was unable to classify the perturbed opera icon at all. Actually, When the standard deviation is increased to 5, the network misclassified the perturbed Opera icon as a Firefox icon with a confidence level of 76%. Thus, from these results, we believe that to some extent using Gaussian distributed noise can be an effective and alternative method to defend from AI-powered GUI attack.

Applying salt-and-pepper noise to our image was not quite as successful in blocking the neural network from identifying Google Chrome icons within the image. During our testing of the object detection network using an image generated with salt-and-pepper noise, the neural network was able to distinguish the icon for Google Chrome with an 88-90% confidence level. For Microsoft Edge, the confidence level of recognition hovered at around 95-97%. Mozilla Firefox required a slightly larger amount of noise than other icons to drop the confidence level to 80%. As for the Opera icon, the addition of salt-and-pepper noise was completely ineffective in reducing the confidence level of recognition by the neural network. When adjusting the amount of noise in the image, the Salt-and-Pepper method of noise addition was quite variable. A small edit to the noise function could drastically change the image. If the level was chosen such that the icon was still recognizable, the confidence level of recognition hovered between 80 and 95 percent. With more change to the noise function, the icons were almost unrecognizable to the human eye and because of this, the image classifier was also unable to recognize the icons.

Slight changes to the value of lambda in the Poisson function caused large changes in image perturbation. For example, we used a lambda value of 500 to run this image through the classifier, the chrome icon was unrecognizable to the neural network. Although the chrome icon was still recognizable to humans, it looked so noised. Using a much smaller lambda value of 100, the image was recognized by the neural net with the confidence of 88%, however, the pixel colors were drastically reversed. Microsoft Edge, Firefox icon, and Opera have the similar effect when applying lambdas of 500 and 100 respectively. At this point in time, it showed that Poisson distributed noise was an effective method of fooling our image classifier to some extent but so much noise made the image hard to recognized by human eyes.

After creating and adding the speckle noise to the Google Chrome icon and pasting it back into our background image, the icon was almost completely overpowered by the noise. The shape of the icon was still the same, but was not so recognizable as chrome to the eye in any way. This directly led to the image classifier being unable to recognize the icon as well. In two out of the four icons we tested, speckle noise completely overpowered the image: Google Chrome and Mozilla Firefox. For Microsoft Edge and Opera, they are still recognizable to the human eye after the addition of speckle noise. However, since we want our adversarial icons to still be easily recognized by the human eye for all cases, speckle noise is not a viable option in our experiment to fend off a GUI-Attack from an AI image classifier.

## 7 CONCLUSION

This article has demonstrated a generic and low-cost malware prototype that can perform an effective attack targeting at the popular GUI systems. According to the results in Section 4, for an original image that is not processed or prevented, the recognition model of a GUI malware was able to recognize the objects such as icons on the desktop consistently with a confidence level up to 99% and on the application tray with the confidence level for usually by approximately 2-4 percentage points loss respectively. The high confidence level of recognition performed by the complicated and advanced AI-powered models has underpinned the GUI malware to commit an effective and low-cost GUI attack.

On the other side, despite the complicated and advanced nature of AI models such as neural networks have the potential to be readily fooled by properly constructed adversarial examples. The adversarial examples can prevent the image classifier from identifying the input image at all or even force it to recognize it as something else with a high level of confidence. According to the results in Section 6, the adversarial examples based methods have shown its viability as a countermeasure to fend off the GUI attack by replacing the original GUI components such as icons with perturbed ones. For other noise-adding methods, their performance vary. For some methods such as poisson, salt-pepper, and speckle, although they can make the AI model misclassify the GUI components, they also impact the visionary discretion of human eyes to some extent. Gaussian noise has a bit better performance over others by deceiving the judgment of a learning machine without so much visionary impacts to human eyes.

Overall, we used adversarial examples as an attempt to hide and mask our web browser icons and prevent a neural network from recognizing them while keeping them as close to the original icon as possible, thus protecting our information. The Fast Gradient Sign Method was the most effective in our testing, with the perturbed icon still looking very similar to the original and easily recognized by human eyes after being pasted back into the desktop background. We believe that the experimental results have shown that adversarial examples have great potential as a tool to mitigate the threats of the GUI attack in the very near future.

In this article, we have illustrated the GUI attack, an application of AI-powered image recognition, and its countermeasure, an application of adversarial examples on cybersecurity, despite the prototype stage of GUI malware and the preliminary results of defensive methods. Actually, it also indicates the lack of transparency behind AI-based decision making, which can lead to security threats, unexpected behavior, and even severe consequences [6]. Thus, in the future, on the one hand, we will continue to utilize AI-powered tools to reveal more security threats on various applications; and on the other hand, we will further study the interpretability and explainability of a learning machine to mitigate the security threats from the algorithmic defect of machine learning.